\documentclass[twoside,a4paper,11pt]{sea10}
\usepackage{graphicx}
\usepackage{hyperref}
\usepackage{movie15}
\def\hoy{\number\day \space de \space\ifcase\month\or
 Enero\or Febrero\or Marzo\or Abril\or Mayo\or Junio\or
 Julio\or Agosto\or Septiembre\or Octubre\or Noviembre\or Diciembre\fi
 \space de \number\year}
\def\ii/{\'{\i}}
\def\cion/{ci\'on}
\def\cao/{\c c\~ao}
\def\degr{\hbox{$^\circ$}}

\def\utw{\smash{\rlap{\lower5pt\hbox{$\sim$}}}}
\def\udtw{\smash{\rlap{\lower6pt\hbox{$\approx$}}}}

\def\tens#1{\ifmmode\mathchoice{\mbox{$\sf\displaystyle#1$}}
{\mbox{$\sf\textstyle#1$}}
{\mbox{$\sf\scriptstyle#1$}}
{\mbox{$\sf\scriptscriptstyle#1$}}\else
\hbox{$\sf\textstyle#1$}\fi}
\def\vec#1{\ifmmode\mathchoice{\mbox{\boldmath$\displaystyle#1$}}
{\mbox{\boldmath$\textstyle#1$}}
{\mbox{\boldmath$\scriptstyle#1$}}
{\mbox{\boldmath$\scriptscriptstyle#1$}}\else
\hbox{\boldmath$\textstyle#1$}\fi}
\def\bbbc{{\mathchoice {\setbox0=\hbox{$\displaystyle\rm C$}\hbox{\hbox
to0pt{\kern0.4\wd0\vrule height0.9\ht0\hss}\box0}}
{\setbox0=\hbox{$\textstyle\rm C$}\hbox{\hbox
to0pt{\kern0.4\wd0\vrule height0.9\ht0\hss}\box0}}
{\setbox0=\hbox{$\scriptstyle\rm C$}\hbox{\hbox
to0pt{\kern0.4\wd0\vrule height0.9\ht0\hss}\box0}}
{\setbox0=\hbox{$\scriptscriptstyle\rm C$}\hbox{\hbox
to0pt{\kern0.4\wd0\vrule height0.9\ht0\hss}\box0}}}}
\def\bbbq{{\mathchoice {\setbox0=\hbox{$\displaystyle\rm
Q$}\hbox{\raise
0.15\ht0\hbox to0pt{\kern0.4\wd0\vrule height0.8\ht0\hss}\box0}}
{\setbox0=\hbox{$\textstyle\rm Q$}\hbox{\raise
0.15\ht0\hbox to0pt{\kern0.4\wd0\vrule height0.8\ht0\hss}\box0}}
{\setbox0=\hbox{$\scriptstyle\rm Q$}\hbox{\raise
0.15\ht0\hbox to0pt{\kern0.4\wd0\vrule height0.7\ht0\hss}\box0}}
{\setbox0=\hbox{$\scriptscriptstyle\rm Q$}\hbox{\raise
0.15\ht0\hbox to0pt{\kern0.4\wd0\vrule height0.7\ht0\hss}\box0}}}}
\def\bbbt{{\mathchoice {\setbox0=\hbox{$\displaystyle\rm
T$}\hbox{\hbox to0pt{\kern0.3\wd0\vrule height0.9\ht0\hss}\box0}}
{\setbox0=\hbox{$\textstyle\rm T$}\hbox{\hbox
to0pt{\kern0.3\wd0\vrule height0.9\ht0\hss}\box0}}
{\setbox0=\hbox{$\scriptstyle\rm T$}\hbox{\hbox
to0pt{\kern0.3\wd0\vrule height0.9\ht0\hss}\box0}}
{\setbox0=\hbox{$\scriptscriptstyle\rm T$}\hbox{\hbox
to0pt{\kern0.3\wd0\vrule height0.9\ht0\hss}\box0}}}}
\def\bbbs{{\mathchoice
{\setbox0=\hbox{$\displaystyle     \rm S$}\hbox{\raise0.5\ht0\hbox
to0pt{\kern0.35\wd0\vrule height0.45\ht0\hss}\hbox
to0pt{\kern0.55\wd0\vrule height0.5\ht0\hss}\box0}}
{\setbox0=\hbox{$\textstyle        \rm S$}\hbox{\raise0.5\ht0\hbox
to0pt{\kern0.35\wd0\vrule height0.45\ht0\hss}\hbox
to0pt{\kern0.55\wd0\vrule height0.5\ht0\hss}\box0}}
{\setbox0=\hbox{$\scriptstyle      \rm S$}\hbox{\raise0.5\ht0\hbox
to0pt{\kern0.35\wd0\vrule height0.45\ht0\hss}\raise0.05\ht0\hbox
to0pt{\kern0.5\wd0\vrule height0.45\ht0\hss}\box0}}
{\setbox0=\hbox{$\scriptscriptstyle\rm S$}\hbox{\raise0.5\ht0\hbox
to0pt{\kern0.4\wd0\vrule height0.45\ht0\hss}\raise0.05\ht0\hbox
to0pt{\kern0.55\wd0\vrule height0.45\ht0\hss}\box0}}}}
\def\bbbz{{\mathchoice {\hbox{$\sf\textstyle Z\kern-0.4em Z$}}
{\hbox{$\sf\textstyle Z\kern-0.4em Z$}}
{\hbox{$\sf\scriptstyle Z\kern-0.3em Z$}}
{\hbox{$\sf\scriptscriptstyle Z\kern-0.2em Z$}}}}
\def\diameter{{\ifmmode\mathchoice
{\ooalign{\hfil\hbox{$\displaystyle/$}\hfil\crcr
{\hbox{$\displaystyle\mathchar"20D$}}}}
{\ooalign{\hfil\hbox{$\textstyle/$}\hfil\crcr
{\hbox{$\textstyle\mathchar"20D$}}}}
{\ooalign{\hfil\hbox{$\scriptstyle/$}\hfil\crcr
{\hbox{$\scriptstyle\mathchar"20D$}}}}
{\ooalign{\hfil\hbox{$\scriptscriptstyle/$}\hfil\crcr
{\hbox{$\scriptscriptstyle\mathchar"20D$}}}}
\else{\ooalign{\hfil/\hfil\crcr\mathhexbox20D}}%
\fi}}
\def\sq{\ifmmode\squareforqed\else{\unskip\nobreak\hfil
\penalty50\hskip1em\null\nobreak\hfil\squareforqed
\parfillskip=0pt\finalhyphendemerits=0\endgraf}\fi}
\def\squareforqed{\hbox{\rlap{$\sqcap$}$\sqcup$}}
\newcommand{\nodata}{\ldots}
\begin{document}
\pagenumbering{arabic}
\pagestyle{myheadings}
\thispagestyle{empty}
{\flushleft\includegraphics[width=\textwidth,bb=58 650 590 680]{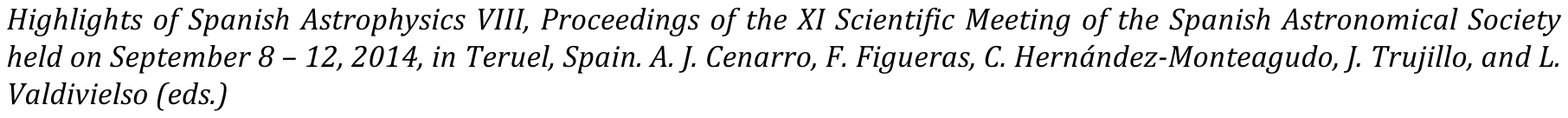}}
\vspace*{0.2cm}
\begin{flushleft}
{\bf {\LARGE
%
%%% TITLE of the paper. 
%%% TITLE of the paper. 
MGB and the new Galactic O-Star Spectroscopic Survey spectral classification standard grid
%
% Do not delete next few lines
}\\
\vspace*{1cm}
%
%%% Include here the LIST OF AUTHORS.
%%% Include here the LIST OF AUTHORS.
%%% Note that the last author has to be preceeded by an AND.
J. Ma{\'\i}z Apell{\'a}niz$^1$, 
E. J. Alfaro$^2$,
J. I. Arias$^3$,
R. H. Barb{\'a}$^3$,
R. C. Gamen$^4$,
A. Herrero$^{5,6}$,
J. R. S. Le\~ao$^7$,
A. Marco$^8$,
I. Negueruela$^8$,
S. Sim{\'o}n-D{\'\i}az$^{5,6}$,
A. Sota$^2$,
and 
N. R. Walborn$^9$
%
% Do not delete next few lines
}\\
\vspace*{0.5cm}
%
%%% AFFILIATIONS LIST.
%%% and the AFFILIATIONS LIST. Note that one affiliation per line.
%%% Add as many affiliations as necessary. 
$^1$
Centro de Astrobiolog{\'\i}a, INTA-CSIC, Spain\\
$^2$
Instituto de Astrof{\'\i}sica de Andaluc{\'\i}a-CSIC, Spain\\
$^3$
Universidad de La Serena, Chile\\
$^4$
Instituto de Astrof{\'\i}sica de La Plata, Argentina\\
$^5$
Instituto de Astrof{\'\i}sica de Canarias, Spain\\
$^6$
Universidad de La Laguna, Spain\\
$^7$
Universidade Federal do Rio Grande, Brazil\\
$^8$
Universidad de Alicante, Spain\\
$^9$
Space Telescope Science Institute, USA
%
% Do not delete next few lines
\end{flushleft}
%
% Headings
\markboth{
%%% Type the SHORT version of the paper title.
%%% Type the SHORT version of the paper title.
MGB and the new GOSSS spectral classification standard grid
}{ % Do not delete
%
%%%  First Author \& Second Author   OR   First-author et al. 
%%%  First Author \& Second Author   OR   First-author et al. if the author list 
%%% contains three or more authors.
Ma{\'\i}z Apell{\'a}niz et al.
% 
% Do not delete next few lines
}
\thispagestyle{empty}
\vspace*{0.4cm}
\begin{minipage}[l]{0.09\textwidth}
\ 
\end{minipage}
\begin{minipage}[r]{0.9\textwidth}
\vspace{1cm}
\section*{Abstract}{\small
%
% ABSTRACT ABSTRACT ABSTRACT
% ABSTRACT ABSTRACT ABSTRACT
%%% Type the ABSTRACT of your paper
In this poster we present three developments related to the Galactic O-Star Spectroscopic Survey 
(\href{http://adsabs.harvard.edu/abs/2011hsa6.conf..467M}{GOSSS}). First, we 
are making public the first version of MGB, an IDL code that allows the user to compare oberved spectra to a grid of spectroscopic 
standards to measure spectral types, luminosity classes, rotation indexes, and spectral qualifiers. Second, we present the 
associated grid of standard stars for the spectral types O2 to O9.7, with several improvements over the original GOSSS grid of 
\href{http://adsabs.harvard.edu/abs/2011ApJS..193...24S}{Sota et al. (2011)}.
Third, we present a list of egregious classification errors in \href{http://simbad.u-strasbg.fr/simbad/}{SIMBAD}: stars that are or have been 
listed there as being of O type but that in reality are late-type stars.
%
% Do not delete next few lines
\normalsize}
\end{minipage}

\begin{figure}
%\centerline{\includegraphics*[width=1.0\linewidth, bb=0 0 1728 792]{HD_93_161_A_B2500_inv.pdf}}
\centerline{\includegraphics*[width=1.0\linewidth]{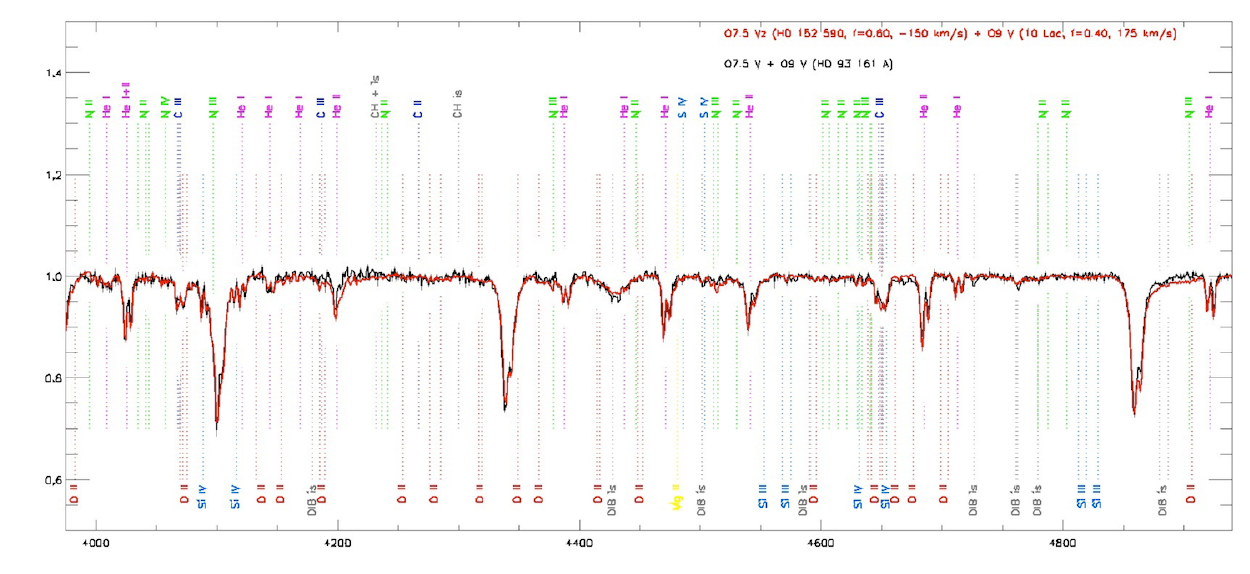}}
\caption{Example of fitting an SB2 system with MGB. Eight parameters can be adjusted: the spectral subtypes, 
luminosity classes, and velocities of both the primary and secondary, the flux fraction of the secondary, and 
the rotation index n. Here HD~93\,161~A (black) is fitted with a combination (red) of 60\% of HD~152\,590 and 
40\% of 10~Lac separated by 325 km/s.}
\label{fig1}
\end{figure}

\section{What is GOSSS?}

$\,\!$\indent GOSSS stands for Galactic O-Star Spectroscopic Survey (\href{http://adsabs.harvard.edu/abs/2011hsa6.conf..467M}{Ma{\'\i}z Apell\'aniz et al. 2011}).
In this project we are observing all Galactic stars that anybody has ever classified as O (if we get time on a large enough telescope)
with $R\sim$ 2500 spectroscopy in the blue-violet region and 
a S/N $\sim$ 300 (in $\sim$90\% of the cases).
The telescopes used so far are: 1.5 m OSN, 3.5 m CAHA, WHT, and GTC (north); 2.5 m LCO and Gemini (south).
We have 2000+ stars observed so far, with completeness to $B=8$ ($B=10$ by 2015) and objects as dim as $B=16$ ($B=19$ planned).
In some cases we have multiple epochs for extreme SB2s and variables.
GOSSS uses a devoted pipeline and quality control systems.

\begin{table}
\caption{The OB2500 v2.0 grid of standards.}
\tiny  
\centerline{
\begin{tabular}{llllllll}
\\
\hline
 & \multicolumn{1}{c}{V} & \multicolumn{1}{c}{IV} & \multicolumn{1}{c}{III} & \multicolumn{1}{c}{II} & \multicolumn{1}{c}{Ib} & \multicolumn{1}{c}{Iab/I} & \multicolumn{1}{c}{Ia} \\
\hline
O2   &                       &                      &                        &                         &                   & {\it HD 93\,129 AaAb} &                    \\
\hline
O3   & {\it HD 64\,568}      &                      & \nodata                &                         &                   &      Cyg OB2-7        &                    \\
\hline
O3.5 & {\it HD 93\,128}      &                      & {\it Pismis 24-17}     &                         &                   & \nodata               &                    \\
\hline
O4   & {\bf HD 46\,223}      &                      & {\bf HD 168\,076 AB}   &                         &                   &      HD 15\,570       &                    \\
     & {\it HD 96\,715}      &                      & {\it HD 93\,250 AB}    &                         &                   &      HD 16\,691       &                    \\
     &                       &                      &                        &                         &                   &      HD 190\,429 A    &                    \\
\hline
O4.5 &      HD 15\,629       &                      & \nodata                &                         &                   &      HD 14\,947       &                    \\
     & {\it HDE 303\,308 AB} &                      &                        &                         &                   &      Cyg OB2-9        &                    \\
\hline
O5   & {\it HDE 319\,699}    &                      & {\bf HD 168\,112}      &                         &                   & {\it CPD -47 2963}    &                    \\
     & {\bf HD 46\,150}      &                      & {\it HD 93\,843}       &                         &                   &                       &                    \\
\hline
O5.5 & {\it HD 93\,204}      &                      & \nodata                &                         &                   &      Cyg OB2-11       &                    \\
     &                       &                      &                        &                         &                   & {\it ALS 18 747}      &                    \\
\hline
O6   & {\it CPD -59 2600}    & {\it HD 101\,190}    & \nodata                &      HDE 229\,196       & \nodata           & \nodata               & {\bf HD 169\,582}  \\
     &      HD 42\,088       &                      &                        &                         &                   &                       &                    \\
     & {\it HDE 303\,311}    &                      &                        &                         &                   &                       &                    \\
\hline
O6.5 & {\bf HD 167\,633}     & {\it HDE 322\,417}   &      HD 190\,864       & {\bf HD 157\,857}       & \nodata           & \nodata               & {\it HD 163\,758}  \\
     & {\it HD 91\,572}      &                      & {\it HD 96\,946}       &                         &                   &                       &                    \\
     &      HD 12\,993       &                      & {\it HD 152\,723 AaAb} &                         &                   &                       &                    \\
     &                       &                      & {\it HD 156\,738}      &                         &                   &                       &                    \\
\hline
O7   & {\it HD 93\,146 A}    & \nodata              &      Cyg OB2-4 A       & {\it HD 94\,963}        & {\it HD 69\,464}  & \nodata               & \nodata            \\
     &      HDE 242\,926     &                      & {\it HD 93\,160}       & {\it HD 151\,515}       &      HD 193\,514  &                       &                    \\
     & {\it HD 91\,824}      &                      &                        &                         &                   &                       &                    \\
     & {\it HD 93\,222}      &                      &                        &                         &                   &                       &                    \\
\hline
O7.5 & {\it HD 152\,590}     & \nodata              & {\it HD 163\,800}      &      HD 34\,656         &      HD 17\,603   &      HD 192\,639      & \nodata            \\
     &      HD 35\,619       &                      &                        & {\bf HD 171\,589}       & {\it HD 156\,154} & {\bf 9 Sge}           &                    \\
\hline
O8   & {\it HD 101\,223}     & {\it HD 94\,024}     & {\it HDE 319\,702}     & {\it 63 Oph}            & {\bf BD -11 4586} &      HD 225\,160      & {\it HD 151\,804}  \\
     & {\it HD 97\,848}      & {\it HD 135\,591}    & {\bf $\lambda$ Ori A}  &                         &                   &                       &                    \\
     &      HD 191\,978      &                      &                        &                         &                   &                       &                    \\
\hline
O8.5 & {\it HDE 298\,429}    & {\bf HD 46\,966}     & {\it HD 114\,737 AB}   & {\it HD 75\,211}        & {\it HD 125\,241} & \nodata               & {\it HDE 303\,492} \\
     &      HD 14\,633       &                      &      HD 218\,195 A     &      HD 207\,198        &                   &                       &                    \\
     & {\bf HD 46\,149}      &                      &                        &                         &                   &                       &                    \\
     & {\it HD 57\,236}      &                      &                        &                         &                   &                       &                    \\
     & {\it Trumpler 14-9}   &                      &                        &                         &                   &                       &                    \\
\hline
O9   &      10 Lac           & {\it HD 93\,028}     & {\it HD 93\,249 A}     & {\it HD 71\,304}        &      19 Cep       &      HD 202\,124      &      $\alpha$ Cam  \\
     &      HD 216\,898      & {\it CPD -41 7733}   &      HD 24\,431        & {\it $\tau$ CMa AaAb}   &                   & {\it HD 152\,249}     &                    \\
     & {\it CPD -59 2551}    &                      &      HD 193\,443 AB    &                         &                   &      HD 210\,809      &                    \\
\hline
O9.2 & {\bf HD 46\,202}      & {\it HD 96\,622}     & {\it CPD -35 2105 AB}  & \nodata                 & {\it HD 76\,968}  & {\it HD 154\,368}     & {\it HD 152\,424}  \\
     &      HD 12\,323       &                      &      HD 16\,832        &                         &                   & {\it HD 123\,008}     &                    \\
     &                       &                      &                        &                         &                   &      HD 218\,915      &                    \\
\hline
O9.5 &      AE Aur           &      HD 192\,001     & {\it HD 96\,264}       & {\bf $\delta$ Ori AaAb} & \nodata           &      HD 188\,209      & \nodata            \\
     & {\it $\mu$ Col}       & {\it HD 93\,027}     &                        &                         &                   &                       &                    \\
     &                       & {\it HD 155\,889 AB} &                        &                         &                   &                       &                    \\
\hline
O9.7 & {\bf $\upsilon$ Ori}  &      HD 207\,538     &      HD 189\,957       & {\it HD 68\,450}        & {\bf HD 47\,432}  &      HD 225\,146      &      HD 195\,592   \\
     &                       &                      & {\it HD 154\,643}      & {\it HD 152\,405}       & {\it HD 154\,811} & {\it $\mu$ Nor}       & {\it GS Mus}       \\
     &                       &                      &                        &      HD 10\,125         & {\it HD 152\,147} & {\it HD 104\,565}     &                    \\
     &                       &                      &                        &                         &                   &      HD 191\,781      &                    \\
\hline
Notes & \multicolumn{7}{l}{Normal, {\it italic}, and {\bf bold} typefaces are used for stars with $\delta > +20\degr$, $\delta < -20\degr$, and the equatorial intermediate region, respectively.}
\end{tabular}

\label{tab1}
}
\end{table}

\section{GOSSS goals}

$\,\!$\indent The primary goal of GOSSS is spectral classification. More specifically, we aim to:

\begin{itemize}
 \item Identify and classify all optically accessible Galactic O stars.
 \item Improve classification criteria and possibly define new special types.
 \item Identify objects wrongly classified as O.
\end{itemize}

GOSSS also has five secondary goals:

\begin{itemize}
 \item Derive physical properties of O stars.
 \item Study SB2s in collaboration with high-resolution sister surveys (OWN, CAF\'E-BEANS, IACOB, and NoMaDS, see contributions by 
        I. Negueruela and S. Sim\'on-D{\'\i}az in these proceedings).
 \item Study the extinction law and study its relationship with the ISM (see contribution by J. Ma{\'\i}z Apell\'aniz in these proceedings).
 \item Analyze the spatial distribution of massive stars and dust.
 \item Obtain the massive-star IMF.
\end{itemize}

\begin{table}
\caption{Stars classified as O in \href{http://simbad.u-strasbg.fr/simbad/}{SIMBAD} that are actually of spectral types A to K.}
\small
\centerline{$\,\!$}
\centerline{
\begin{tabular}{lllll} 
\hline
Name & \multicolumn{2}{c}{Spectral type} & SIMBAD reference                                                                 & Notes                                 \\
                    & SIMBAD & New &                                                                                        &                                       \\
\hline
BD -03 2178         & O5      & K & \href{http://adsabs.harvard.edu/abs/1976AJ.....81..225M}{MacConnell \&}                 & Recently fixed in SIMBAD, confusion   \\
                    &         &   & $\;\;$\href{http://adsabs.harvard.edu/abs/1976AJ.....81..225M}{Bidelman (1976)}         & $\;\;$with BD -03 2179, a sdO         \\
BD +01 3974         & O       & F & \href{http://adsabs.harvard.edu/abs/1986SAAOC..10...27K}{Kelly \& Kilkenny (1986)}      &                                       \\
BD +32 4642 A       & O       & F & Not given$^*$                                                                           &                                       \\
BD +37 3929         & O8f     & F & \href{http://adsabs.harvard.edu/abs/1956ApJ...124..367H}{Hiltner \& Johnson (1956)}     & Confusion with BD +37 3927            \\
BD +40 4213         & O9.5 I  & F & \href{http://adsabs.harvard.edu/abs/1991AJ....101.1408M}{Massey \&}                     & Not in the original reference, likely \\
                    &         &   & $\;\;$\href{http://adsabs.harvard.edu/abs/1991AJ....101.1408M}{Thompson (1991)}         & $\;\;$transcription error in SIMBAD   \\
BD +45 4132 A       & O       & F & Not given$^*$                                                                           &                                       \\
BD +61 100 AB       & O/B2    & G & \href{http://adsabs.harvard.edu/abs/1989AbaOB..66...33R}{Radoslavova (1989)}            &                                       \\
CPD $-$61 4623      & O       & K & Not given$^*$                                                                           &                                       \\
HDE 226\,144        & O9 V    & A & \href{http://adsabs.harvard.edu/abs/1980AcA....30..347M}{Mikolajewska \&}               &                                       \\
                    &         &   & $\;\;$\href{http://adsabs.harvard.edu/abs/1980AcA....30..347M}{Mikolajewski (1980)}     &                                       \\
Tyc 0468-02112-1    & O\ldots & F & Not given$^*$                                                                           &                                       \\
\hline
\end{tabular}
}
\medskip

$^*$ These classifications were removed from \href{http://simbad.u-strasbg.fr/simbad/}{SIMBAD} after this poster was presented.
\label{tab2}
\end{table}

\section{MGB}

$\,\!$\indent MGB is a code that attacks spectral classification (\href{http://adsabs.harvard.edu/abs/2012ASPC..465..484M}{Ma{\'\i}z Apell\'aniz et al. 2012})
by doing classical visual (non-automatic) spectral classification by interactively comparing with a standard grid.
The MGB user can adjust four parameters:

\begin{itemize}
 \item Spectral subtype (horizontal classification).
 \item Luminosity class (vertical classification).
 \item n index (broadening).
 \item Alternative standards at each grid point (e.g. ONC or f variants).
\end{itemize}

MGB also includes fitting of SB2 systems (Figure~\ref{fig1}).
The default grid covers the O2-O9.7 spectral subtypes using GOSSS data (see below).
Other grids (O-type or other) at various resolutions using the original or degraded spectra from different on-going high-resolution surveys (e.g. IACOB, OWN, IACOBsweG) are planed.
MGB v1.0 is available now from \url{http://jmaiz.iaa.es}.

\begin{figure}
\centerline{\includegraphics*[width=1.00\linewidth, bb=28 28 566 674]{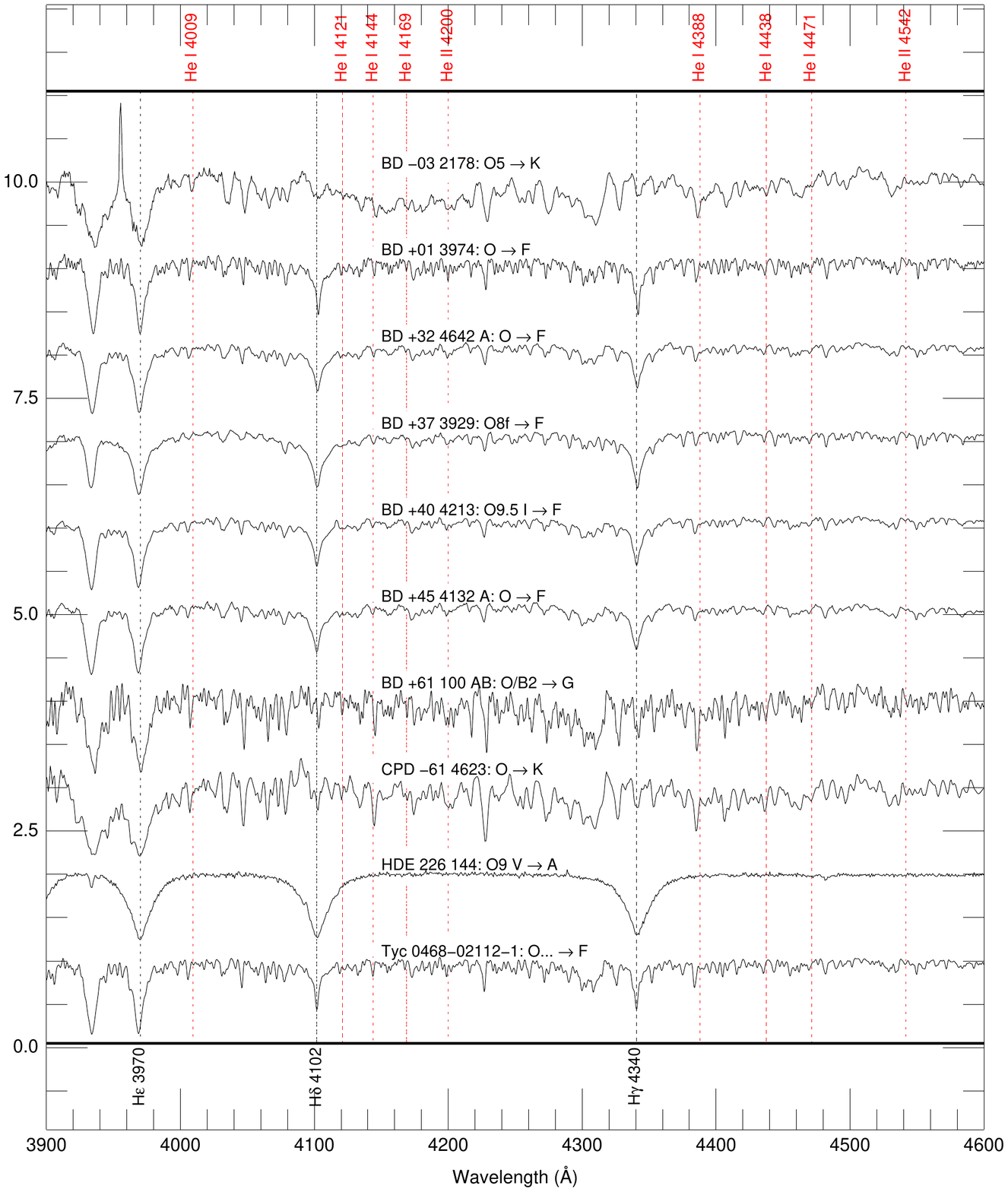}}
\caption{Spectrograms for the stars in Table~\ref{tab2}.}
\label{fig2}
\end{figure}

\section{The new GOSSS standard grid}

$\,\!$\indent We present the OB2500 v2.0 GOSSS standard grid, which is integrated with MGB.
It covers the spectral subtypes from O2 to O9.7 and the luminosity classes from V to Ia (Table~\ref{tab1}).
The grid has two types of gaps: non-existing types (blank) and standards not yet found (\ldots).
It is similar to OB2500 v1.0, the grid in \href{http://adsabs.harvard.edu/abs/2011ApJS..193...24S}{Sota et al. (2011)}, 
        but with some small changes introduced by \href{http://adsabs.harvard.edu/abs/2014ApJS..211...10S}{Sota et al. (2014)} 
        e.g. the addition of O9.2 and new standards.
The grid is available from \url{http://jmaiz.iaa.es} with MGB v1.0.
A future extension to A0 (including all B stars) and luminosity class Ia+ is planned.

\section{Spectral classification errors}

$\,\!$\indent During the course of GOSSS we have discovered a large number of classification errors in the literature. More specifically:

\begin{itemize}
 \item 24.9\% of the alleged O stars observed by mid 2013 were not of that type (false positives, 
        \href{http://adsabs.harvard.edu/abs/2013msao.confE.198M}{Ma{\'\i}z Apell\'aniz et al. 2013}).
 \item The current number of false positives is closer to 30\%.
 \item False negatives are much lower (6.4\%, \href{http://adsabs.harvard.edu/abs/2013msao.confE.198M}{Ma{\'\i}z Apell\'aniz et al. 2013}).
\end{itemize}

\href{http://simbad.u-strasbg.fr/simbad/}{SIMBAD} has many errors in O-type spectral classifications, which are related to different issues:

\begin{itemize}
 \item Some spectral types are actually of photometric, not spectroscopic origin.
 \item Other classifications are of unknown origin (no reference is provided).
 \item Misidentifications (in the source or in \href{http://simbad.u-strasbg.fr/simbad/}{SIMBAD}) are present.
 \item Sometimes the lower quality classification is shown at the top, leaving the higher quality one ``hidden'' in the text below.
% \item It is difficult to have (even egregious) errors corrected.
\end{itemize}

The most egregious errors we have found are A-K stars that appear or have appeared in \href{http://simbad.u-strasbg.fr/simbad/}{SIMBAD} as O stars 
(Table~\ref{tab2} and Figure~\ref{fig2}).

\section*{References}
\begin{itemize}
 \item \href{http://adsabs.harvard.edu/abs/1956ApJ...124..367H}{Hiltner, W. A. \& Johnson, H. L. 1956, {\it ApJ} {\bf 124}, 367}.
 \item \href{http://adsabs.harvard.edu/abs/1986SAAOC..10...27K}{Kelly, B. D. \& Kilkenny, D. 1986, {\it SAAOC} {\bf 10}, 27}.
 \item \href{http://adsabs.harvard.edu/abs/1976AJ.....81..225M}{MacConnell, D. J. \& Bidelman, W. P. 1976, {\it AJ} {\bf 81}, 225}.
 \item \href{http://adsabs.harvard.edu/abs/2011hsa6.conf..467M}{Ma{\'{\i}}z Apell{\'a}niz, J. et al. 2011, {\it Highlights of Spanish Astrophysics VI}, 467}.
 \item \href{http://adsabs.harvard.edu/abs/2012ASPC..465..484M}{Ma{\'{\i}}z Apell{\'a}niz, J. et al. 2012, {\it ASP conference series} {\bf 465}, 484}.
 \item \href{http://adsabs.harvard.edu/abs/2013msao.confE.198M}{Ma{\'{\i}}z Apell{\'a}niz, J. et al. 2013, {\it Massive Stars: From $\alpha$ to $\Omega$}, 198}.
 \item \href{http://adsabs.harvard.edu/abs/1991AJ....101.1408M}{Massey, P. \& Thompson, A.~B. 1991, {\it AJ} {\bf 101}, 1408}.
 \item \href{http://adsabs.harvard.edu/abs/1980AcA....30..347M}{Mikolajewska, J. \& Mikolajewski, M. 1980, {\it AcA} {\bf 30}, 347}.
 \item \href{http://adsabs.harvard.edu/abs/1989AbaOB..66...33R}{Radoslavova, T. 1989, {\it AbaOB} {\bf 66}, 33}.
 \item \href{http://adsabs.harvard.edu/abs/2011ApJS..193...24S}{Sota, A. et al. 2011, {\it ApJS} {\bf 193}, 24}.
 \item \href{http://adsabs.harvard.edu/abs/2014ApJS..211...10S}{Sota, A. et al. 2014, {\it ApJS} {\bf 211}, 10}.
\end{itemize}

\end{document}